\documentstyle[12pt,aaspp4]{article}
\input epsf
\def\pref#1{(\ref{#1})}
\def\ep{\epsilon}
\def\ren{R_{m}}
\def\nn{\nonumber \\ }
\def\pref#1{(\ref{#1})}

\def\ni{\noindent}
\def\lra#1{\left\langle #1\right\rangle}

\def\bar{\overline}
\def\OV{\overline{\bf V}} 
\def\OB{\overline{\bf B}}
\def\emf{\overline{\mbox{${\cal E}$}} {}}
\def\emfb{\overline{\mbox{\boldmath ${\cal E}$}} {}}
\def\bbE{\bar {\bf E}}

\def\beq{\begin{equation}}
\def\ee{\end{equation}}
\def\lsim{\mathrel{\rlap{\lower4pt\hbox{\hskip1pt$\sim$}}
    \raise1pt\hbox{$<$}}}
\def\gsim{\mathrel{\rlap{\lower4pt\hbox{\hskip1pt$\sim$}}
    \raise1pt\hbox{$>$}}}
\def\bfE{{\bf E}}

\def\bfA{{\bf A}}
\def\bfa{{\bf a}}
\def\bfe{{\bf e}}
\def\bfB{{\bf B}}
\def\bbJ{\bar {\bf J}}

\def\bB{\overline B}
\def\ts{\times}
\def\lb{\langle}
\def\rb{\rangle}
\def\curl{\nabla {\ts}}
\def\nt{\nabla\times}
\def\cnt{\cdot\nt}

\def\bbV{\bar {\bf V}}
\def\bfv{{\bf v}}

\def\bfj{{\bf j}}

\def\bfe{{\bf e}}

\def\bfb{{\bf b}}

\def\bfB{{\bf B}}

\def\bbB{\overline {\bf B}}
\def\bbA{\overline {\bf A}}

\begin{document}

\setcounter{equation}{0}

\title{Understanding Helical Magnetic Dynamo Spectra 
with a  Nonlinear Four-Scale  Theory}
\medskip

\author{Eric G. Blackman} 
\affil{ Department of Physics \&Astronomy and Laboratory for
Laser Energetics, University of Rochester, Rochester NY 14627;
email:  blackman@pas.rochester.edu}

\centerline {(submitted to  MNRAS)} 

\begin{abstract}

Recent MHD dynamo simulations for magnetic Prandtl number $>1$ 
demonstrate that when MHD turbulence is forced with sufficient kinetic 
helicity, the saturated magnetic energy spectrum evolves from 
having a single peak below the forcing  scale to become  
doubly peaked with one peak at the system (=largest) 
scale and one at the forcing scale. 
The system scale field growth is well modeled by a 
recent nonlinear two-scale nonlinear helical dynamo theory in which 
the system and forcing scales carry magnetic helicity of opposite sign.  
But a two-scale theory cannot model the shift of the 
small-scale peak toward the forcing scale.
Here I develop a  four-scale helical dynamo theory which shows that  
the small-scale helical magnetic energy first saturates at 
very small scales, but then successively saturates at larger values 
at larger scales, eventually becoming dominated by the forcing scale. 
The transfer of the small scale peak  to the forcing scale
is completed by the end of the kinematic growth
regime of the large scale field, 
and does not depend on magnetic Reynolds number $R_M$ for large $R_M$.
The four-scale and two-scale theories subsequently evolve almost identically,  
and both show significant field growth on the system and forcing
scales that is independent of $R_M$.
In the present approach, the helical and nonhelical parts of the spectrum
are largely decoupled. Implications for fractionally helical turbulence
are discussed.


\end{abstract}




\section{Introduction}

Most all ionized astrophysical rotators 
are turbulent and carry magnetic fields. 
For astrophysical objects such as the sun, 
helical dynamo theory was original proposed as a theory to model the 
large scale field generation on scales larger than that of the
driving turbulence (e.g. Parker 1955; Steenbeck, Krause, R\"adler 1966; 
Moffatt 1978; Parker 1979). 
Also of interest is to understand the 
{\it overall} shape and amplitude of the magnetic energy spectrum from
helical or nonhelical dynamo action on both large and small scales
(e.g. \cite{pfl,kulsrud}, Brandenburg et al. 1995; Hawley, Gammie, Balbus 1995,1996; Brandenburg 2001, Shekochihin et al. 2002ab, Maron \& Cowley 2002;
Maron \& Blackman 2002).
When the effect of the growing magnetic field
on the turbulent velocity is ignored, the dynamo model is kinematic.  
We know that kinematic theory is incomplete
since the dynamo must represent a solution of the full nonlinear
magnetohydrodynamic (MHD) equations. In this context, much of
theoretical dynamo research can be divided into two categories: 
(1) that which focuses on the fundamental principles, with the goal
of determining a correct and useful nonlinear theory, and  
(2) that which parameterizes the backreaction of the growing
magnetic field on the turbulent velocities and uses  
it in the dynamo equations to model 
the fields of specific objects in detail.  
This paper falls into the former category.

Turbulence that produces astrophysical dynamos can either
arise from external forcing or be 
self-driven from shear. Examples of the former
are supernova driving in the Galaxy (e.g. Ruzmaikin, Shukurov, Sokoloff 1988)
or thermal convection driving in the sun (e.g. Parker 1979), 
and an example of the latter is the magneto-rotational instability (MRI)
(Balbus \& Hawley 1991;1998).
In either forced or self-driven dynamos, there exist both helical or
 nonhelical varieties.
Nonhelical vs. helical refers to the situation  in which turbulence
has either zero vs. finite  pseudoscalar quantities 
such as the kinetic helicity $\overline{\bfv\cdot\curl\bfv}$, 
where $\bfv$ is the turbulent velocity, and the overbar represents
some volume average (within a given hemisphere for a rotator).
Such a correlation can be derived from the statistical tendency for
rising (falling) eddies to twist in the opposite (same) sense as the underlying
rotation as they expand (contract) and 
conserve angular momentum in stratified rotator.
When kinetic helicity is present, the turbulence can generate a global
large scale field (e.g. Moffatt 1978; Parker 1979)
whose associated sign of flux 
lasts many overturn times of the largest turbulent eddy.  
As discussed in Blackman (2002b),  such a global field is not to be confused 
with fields produced in nonhelical simulations 
for which the largest turbulent
scale is the box scale. In this latter case, magnetic energy will
arise on the scale of the box, but the sign of its flux will change
every overturn time scale.  For example, nonstratified
MRI simulations (Hawley, Gammie, Balbus 1995) 
for which no kinetic helicity is present, 
produce magnetic energy with maximum azimuthal scale of order the box height, 
but the turbulence also extends to this scale, and the field
does not maintain a coherent flux longer than a large eddy turnover time
($\sim$ an orbit time). 
In contrast, helical MRI simulations in a non-periodic box
(Brandenburg et al. 1995) were shown to 
incur a helical dynamo which sustains a large scale 
flux over many orbit times.  
This latter type of global field is extremely helpful
in producing jets and coronae, since it can survive the buoyant
rise without shredding on its way up (Blackman 2003).
(Note that Stone et al. (1996) claimed that no helicity was observed
in stratified periodic box disk simulations. But they averaged over their 
entire box to test for helicity, and since the northen and southern
hemispheres must produce opposite sign, the box averaged helicity would
be expected to vanish.)

As the amplification of magnetic fields and the associated MHD
turbulence  represent highly nonlinear problems, 
understanding the backreaction of the magnetic
field on the field growth itself has been an evolving
topic of study  
(e.g. \cite{piddington,kleeorin82,vc,ch,kulsrud,fbc,bf2000,b2001,fb02,bf02}).
Some recent developments (Brandenburg 2001; Field \& Blackman 2002; 
Blackman \& Field 2002;
Brandenburg, Dobler, Subramanian 2002; Blackman \& Brandenburg 2002; Maron \& Blackman 2002) 
have emerged from focusing numerically 
and analytically 
on the simplest dynamo forced dynamos 
for which the underlying principles can be identified.  
Experiments have been performed 
in which a turbulent velocity spectrum is first established 
in the absence of magnetic fields in a periodic box and then a weak 
seed spectrum of magnetic energy is input (e.g. Meneguzzi, Frisch, Pouquet
1981; Kida, Yanase, Mizushima 1991;  Brandenburg 2001, Maron \& Cowley 2002, Maron \& Blackman 2002).  The nonlinear evolution the magnetic 
energy spectrum to saturation can then be studied as the system is driven.
(Note that these simulations are distinct from those  
in which an initial uniform background field is also assumed e.g. 
Maron \& Goldreich 2002; Cho, Lazarian \& Vishniac 2002)

The shape of the resulting spectrum depends on
the fractional helicity.  Define the fractional kinetic  helicity 
$f_h\equiv |\lb\bfv_f\cdot\curl {\bfv}_f\rb/v_f^2k_2|$, where $\bfv_f$ is the 
turbulent velocity on the forcing scale and $k_f$ is the forcing wavenumber. 
When forced such that $f_h=0$, 
the magnetic spectrum saturates
with the magnetic energy seemingly piling up
close to the resistive scale for magnetic Prandtl number  
(the ratio of viscosity to magnetic diffusivity $Pr_M\equiv\nu/\lambda$)
$>1$, not near the forcing scale (Kida, Yanase, Mizushima 1991; 
Maron \& Cowley 2001;
Schekochihin 2002ab).  This contradicts, for example, 
observations of the Galactic
magnetic field, which does not seem to have a peak on the resistive scale,
(Beck et al. \& 1996).  On the other hand, 
recent simulations of Haugen et al. (2003)
suggest that the peak may not be at the resistive scale even for $Pr>1$.
Determing the location of the peak for the nonhelical dynamo 
for $Pr>1$ is presently an active area of research.
All studies seem to at least agree that the peak is below the forcing scale.

By contrast, simulations that 
force with $f_h \sim 1$ show production of large scale fields
at the box scale and a peak at the forcing scale (Brandenburg 2001).
  Maron and Blackman (2002) studied what 
happens to the spectrum both above and below the forcing
scale as a function of $f_h$. 
By starting with an initial spectrum that represented the saturated state of 
an $f_h=0$ simulation,
they found that for $f_h$ above a critical value ($=k_1/k_2$, where $k_1$ is
the box scale)  a peak at $k_1$ emerged, and as $f_h$ was further
increased toward 1, the peak originally at the sub-forcing scale 
increasingly depleted.  
A peak at the forcing scale emerged.
The adjusted shape of the spectrum at and below the forcing
scale established itself on a kinematic growth time scale of the
$k_1$ field. 
 
The growth rate and saturation value of the large scale field 
at $k_1$ is now well understood 
by a nonlinear two-scale dynamo theory based on magnetic
helicity conservation and exchange between the small scale (assumed to
be the forcing scale) 
and the box (large) scale (Field \& Blackman 2002; Blackman \& Field 2002):  
Growth of the large scale field 
corresponds to a growth of large scale magnetic helicity. 
Magnetic helicity conservation dictates that a small scale helical 
field must then also grow of the opposite sign.
Because the growth driver for the large scale helicity  
also depends on the small scale magnetic helicity, the growth of the
latter ends up quenching the large scale dynamo.
But there is an important limitation of the two-scale theory.
Why should the growth of the small scale compensating magnetic
helicity occur at the forcing scale and not at some much smaller
scale? Answering this can help to understand 
why the small-scale magnetic energy peak initially 
at sub-forcing scales  in simulations of Maron and Blackman (2002)
migrates to the forcing scale when sufficient kinetic helicity
is input. 

To address these questions, I
develop a simple four-scale approach which predicts that 
the small scale magnetic helicity and current helicity
grow fastest on the smallest scale on which there is both a finite
turbulent velocity and magnetic field. (This would be the viscous
scale for $Pr_M\ge 1$ and the resistive scale for $Pr_M \le 1$.)
The magnetic engergy and current helicities 
saturate fastest at these small scales. 
Eventually however, larger values are reached
at larger scales  until the small scale growth 
is dominated by the forcing scale, essentially justifying a 
two-scale approach by 
the time the kinematic growth phase  for the $k_1$ field ends.
It is important to emphasize that the mean field formalism 
herein describes the dynamical evolution of both the large and
small scale magnetic and current 
helicities. The backreaction of the small scale current helicity
on the kinetic helicity, and thus on the growth of the large scale
field is treated dynamically.  We will see that significant 
growth of the large scale field proceeds unimpeded at early times, but
at late times, the backreaction from the small scale field 
slows the growth to a resistively limited pace.  It is imporant
to emphasize that the dynamical mean field theory studied herein, 
albeit a simplified theory, 
goes beyond the kinematic mean field theory of standard textbooks 
which does not include the backreaction.

In section 2, the basic role of helicity conservation in large scale dynamo 
theory is discussed and the four-scale set of equations (3 of which
are coupled) are derived. 
In section 3, the equations are solved. The solutions and their 
interpretation, discussed in section 4. Section 5 is the conclusion.

\section{The Four-scale Nonlinear Dynamo}

The mean field dynamo (MFD) theory has 
been a useful framework for modeling
the in situ origin of large-scale magnetic field 
growth in planets, stars, and galaxies 
(Moffatt 1978; Parker 1979 Krause \& R\"adler 1980; Zeldovich, Ruzmaikin,
Sokoloff 1983)
and has also been invoked 
and to explain the sustenance of fields in  fusion devices
(see Ortolani \& Schnack 1993; Bellan 2000; Ji \& Prager 2002 for reviews). 
Pouquet, Frisch \& Leorat (1976)
derived approximate evolution equations for the spectra of kinetic
energy, magnetic energy, kinetic helicity, and magnetic helicity
and demonstrated the need to consider magnetic helicity in
more carefully in the context of dynamo theory. 
Recent progress in modeling the fully 
nonlinear dynamical field growth and saturation seen in simulations
(Brandenburg 2001) 
has come from dynamically incorporating  magnetic helicity evolution 
using a two-scale theory 
(Field \& Blackman 2002, Blackman \& Brandenburg 2002, Blackman
\& Field 2002).  
Previous attempts (\cite{zeldovich,kleeorin82,kleeorinruz,gd2,by})
to use magnetic helicity in dynamo theory did not produce
a time dependent dynamical theory.




To develop the new four scale theory and make contact
with the previous work, I first average the magnetic 
induction equation to obtain the basic MFD equation
(Moffatt 1978)
\beq
\partial_t\OB= \curl \emfb +\curl(\bbV\ts \bbB) -\lambda \nabla^2\OB,
\label{1}
\ee
where $\OB$ is the mean (large-scale) magnetic field in Alfv\'en speed units,
$ 
\lambda = {\eta c^2\over 4\pi}
$
is the magnetic diffusivity in terms of the resistivity $\eta$, $\bbV$ is
the mean velocity which I set $=0$, and 
$\emfb=\lb\bfv\ts\bfb\rb$
is the turbulent electromotive force, a correlation between
fluctuating velocity $\bfv$ and magnetic field $\bfb$ in Alfv\'en units.  
Instead of the assuming 
that there are two scales as in Field \& Blackman (2002) and 
Blackman \& Field (2002), here I generalize and assume that there
are four scales on which
magnetic helicity can reside, such that $k_4> k_3> k_2>k_1>0$.
An example is to take  $k_2=k_f$, the forcing wavenumber, 
$k_4=k_\lambda$, the resistive wavenumber,
and $k_3=k_\nu$ viscous wavenumber.  
We can interpret $\bbB$, $\bbA$ and $\emfb$ 
as the mean components of $\bfB, \bfA$ and $(\bfv\ts\bfb)$ 
that varies on a scale $k_1^{-1}$ in a closed system.

The total magnetic helicity, $H=\lb\bfA\cdot\bfB\rb_{vol}$, 
satisfies (Woltjer 1958; Moffatt 1978 Berger \& Field 1984)
\beq
\partial_t H=-2\lb{\bfE}\cdot\bfB\rb_{vol},
\label{helcons}
\ee
where $\bf E= -\partial_t {\bf A} -\nabla \phi$,
$\phi$ is the scalar potential, and 
$\lb\rb_{vol}$ indicates a global volume average.
Such a conservation formula  also 
applies separately to each scale. That is, 
using $\bbE = \partial_t \bbA -\nabla{\overline \phi}$, 
$\bfe_2 = \partial_t \bfa_2 -\nabla\phi_2$
and $
\bfe_3= \partial_t \bfa_3 -\nabla\phi_3$
and dotting with $\bbB$, $\bfb_2$ and $\bfb_3$ respectively
we  obtain
\beq
\partial_tH_1=2\lb{\emfb}\cdot\bbB\rb_{vol}
-2\lambda \lb \bbJ\cdot \bbB\rb_{vol},
\label{h1}
\ee
\beq
\partial_tH_2=-2\lb{\bfe_2}\cdot\bfb_2\rb_{vol},
\label{3}
\ee
\beq
\partial_tH_3=-2
\lb{\bfe_3}\cdot\bfb_3\rb_{vol},
\label{4}
\ee
and
\beq
\partial_tH_4=-2
\lb{\bfe_4}\cdot\bfb_4\rb_{vol},
\label{4res}
\ee
where $H_1\equiv\lb\bbA\cdot\bbB\rb_{vol}$, 
$H_2\equiv\lb\bfa_2\cdot\bfb_2\rb_{vol}$, 
 $H_3\equiv\lb\bfa_3\cdot\bfb_3\rb_{vol}$
and $H_4\equiv\lb\bfa_4\cdot\bfb_4\rb_{vol}$.
I now derive expressions for 
$\lb{\emfb}\cdot\bbB\rb_{vol}$,
$\lb{\bfe_2}\cdot\bfb_2\rb_{vol}$, $\lb{\bfe_3}\cdot\bfb_3\rb_{vol}$,
and $\lb{\bfe_4}\cdot\bfb_4\rb_{vol}$. 
I assume that mixed correlations between fluctuating quantities of 
widely separated scales (which fluctuate on widely different time scales)
 do not contribute. Thus I ignore terms
of the form e.g. $\lb\bfb_2\cdot\bfj_3\rb$.
This is an important simplifying
assumption which really needs to be tested numerically,
and I will come back to it again in section 4.
Note however that this assumption is not the same as
ignoring the effect of the small scale fields on the large scale field
growth, or ignoring the nonlinear backreaction
altogether: Eqn. (3-6) along with (8-13) show that  
nonlinear couplings between disparate scales also occur 
because the time evolution for the large scale magnetic helicity
depends non-linearly on the time evolution of the small scale helicities.
These latter couplings are included, even when mixed correlations are
not included.
We then have
\beq
\lb{\emfb}\cdot\bbB\rb_{vol} =
\lb{\emfb}_2\cdot\bbB\rb_{vol} + \lb{\emfb}_3\cdot\bbB\rb_{vol} 
\label{1b}
\ee
where 
$\emfb_2={\overline{\bfv_2\ts\bfb_2}}$
and 
$\emfb_3={\overline{\bfv_3\ts\bfb_3}}$, and 
\beq
-\lb{\bfe_2}\cdot\bfb_2\rb_{vol} =-\lb\emfb_2
\cdot\bbB\rb_{vol}+\lb({\bfv_3\ts\bfb_3})_2\cdot\bfb_2\rb_{vol} -
\lambda\lb \bfj_2\cdot \bfb_2\rb_{vol} 
\label{2b}
\ee
\beq
-\lb{\bfe_3}\cdot\bfb_3\rb_{vol} =
-\lb(\bfv_3\ts\bfb_3)_2\cdot\bfb_2\rb _{vol}
-\lb\emfb_3\cdot\bbB\rb_{vol} 
- \lambda\lb \bfj_3\cdot \bfb_3\rb_{vol} 
\label{3bb}
\ee
and 
\beq
\lb{\bfe_4}\cdot\bfb_4\rb_{vol} = \lambda\lb \bfj_4\cdot \bfb_4\rb_{vol} 
\label{4bb}
\ee
where the overbar indicates averages which vary on scales $k_1^{-1}$
wavenumber, the $()_2$ indicates averages which vary on scales
 $\ge k_2^{-1}$.
Eqn. (\ref{4bb}) follows because $\bfv_4=0$, as I have assumed
$k_4$ is taken at the resistive scale and that mixed scale
correlations  vanish.

In what follows I will solve for the nonlinear time evolution of the magnetic
helicity at each of the four scales. 
Blackman \& Field (2002) showed that a proper nonlinear theory
must technically include the time evolution of $\emfb_2$ and by generalization
here, also the time evolution for $\emfb_3$ into the theory.  However,
they also showed that for  the specific case for which triple correlations
in $\partial_t\emfb_2$ are treated as damping terms with a damping
time $\tau_2=1/k_2v_2$, 
neglecting the time evolution of the turbulent electromotive force
does not qualitatively influence the solution.
Adopting this specific case, 
the implication of their result for the present theory 
is that
\beq
\emfb_2
={\tau_2\over 3} 
(\overline{\bfj_2\cdot\bfb_2}
-\overline{\bfv_2\cdot\curl\bfv_2})\bbB 
-{\overline \beta}_2\bbJ,
\label{8b}
\ee 
where 
$\beta_2$ is a diffusivity computed
from velocities at the $k_2$ scale. (The form of the diffusivity is
not essential for the present discussion, and I will later scale it to its
kinematic value.)
Similarly, we also have 
\beq
\emfb_3=
{\overline{\bfv_3\ts\bfb_3}}=
{\tau_3\over 3} 
({\overline {\bfj_3\cdot\bfb_3}})\bbB
-{\overline \beta}_3\bbJ,
\label{9b}
\ee
and 
\beq
(\bfv_3\ts\bfb_3)_2=
{\tau_3\over 3} 
(\bfj_3\cdot\bfb_3)_2\bfb_2 - (\beta_3)_2\bfj_2,
\label{10b}
\ee
where $\tau_3=1/k_3v_3$, and 
I assume that there are no kinetic helicity 
contributions at the $k_3$ scale. This is reasonable because
I am  focusing on the case in which the 
kinetic helicity is peaked at the forcing scale, 
and kinetic helicity does not cascade efficiently.
If the $k_1$ field is maximally helical,
the current helicity, magnetic helicity and magnetic energy
for the $k_1$ scale are simply related by powers of $k_1$.
(However, the third term of (\ref{2b}) and the second term of (\ref{3bb})
involve a subtly that  will be explained below.) 
Combining equations (6-13) 
we have
\beq
\partial_t H_1=2{\tau_2\over 3}( k_2^2 H_2+f_hk_2 v_2^2) k_1 H_1
+2{\tau_3\over 3}(k_1 H_1 k^2_3H_3)
-2(\lambda +{\overline \beta}_2+{\overline \beta}_3)k_1^2 H_1,
\label{12b}
\ee
\beq
\partial_t H_2=-2{\tau_2\over 3}(k_2^2 H_2+f_hk_2 v_2^2) k_1 H_1
-2{\tau_3\over 3}(k_2 H_2 k^2_3H_3 f_u) +2{\overline \beta}_2k_1^2 H_1
-{2g_u(\beta_3)_2}k_2^2H_2
-2\lambda k_2^2 H_2,
\label{13b}
\ee
\beq
\partial_t H_3=2{\tau_3\over 3}(k_2 H_2 k_3^2H_3 f_u) 
-2{\tau_3\over 3}(k_1 H_1 k^2_3H_3)
+2{\overline \beta}_3k_1^2 H_1
+{2g_u(\beta_3)_2}k_2^2H_2
-2\lambda k_3^2 H_3,
\label{14b}
\ee
and 
\beq
\partial_t H_4=-2\lambda k_4^2 H_4.
\label{14bb}
\ee
The last equation is decoupled from the others,
and represents decay. Thus in present approximation
scheme, where correlations between different
scales vanish and $k_4=k_\lambda \ge  k_3$ there is no
helicity exchange with the resistive scale.  Equation (\ref{14bb})
can then be subsequently ignored.
 
The third and fifth terms of (\ref{13b}) and the second and fifth
terms of (\ref{14b}) involve the quantities $f_u$ and $g_u$.
These are positive quantities and they come from
the subtlety in dotting the second and third terms of 
(\ref{10b})
with $\bfb_2$:  
dotting the third term of (\ref{10b})
with $\bfb_2$ gives the term 
$\propto  \lb(\beta_3)_2\bfj_2\cdot\bfb_2\rb_{vol}=
\lb\beta_3\rb_{vol}\lb
{(\beta_3)_2\over \lb\beta_3\rb_{vol}}
\bfj_2\cdot\bfb_2\rb_{vol}
=\lb\beta_3\rb_{vol}=
\lb{(\beta_3)_2\over \lb\beta_3\rb_{vol}}\bfj_2\cdot\bfb_2\rb_{vol}
=\lb\beta_3\rb_{vol}
g_u k_2^2 H_2$,
where $g_u$ accounts for the deviation of 
${(\beta_3)_2\over\lb\beta_3\rb_{vol}}$ from unity, and I also 
assume that ${\overline {\beta}}_2=\lb\beta_2\rb_{vol}$ and ${\overline {\beta}}_3=\lb\beta_3\rb_{vol}$.  
Similarly, dotting the second term of (\ref{10b})
with $\bfb_2$ and averaging gives a term $\propto$
$\lb(\bfj_3\cdot\bfb_3)_2\bfb_2^2 \rb_{vol}
= 
\lb\bfj_3\cdot\bfb_3\rb_{vol}
\lb{(\bfj_3\cdot\bfb_3)_2\over \lb\bfj_3\cdot\bfb_3\rb_{vol}}
\bfb_2^2 \rb_{vol}\equiv -\lb\bfj_3\cdot\bfb_3\rb_{vol} f_u k_2 H_2$
so that $f_u$ accounts for both the fractional helicity 
at the $k_2$ scale and the deviation of the ratio 
${(\bfj_3\cdot\bfb_3)_2\over \lb\bfj_3\cdot\bfb_3\rb_{vol}}$ 
from unity.  Note that unlike 
$b_2^2$, $b_3^2$ does not enter these equations
so that even if there is a large pile-up in magnetic energy at $k_3$ 
initially, only its force-free component couples into the above equations.

To write equations (\ref{12b}), (\ref{13b}), and (\ref{14b})
in dimensionless form,  I define 
$h_1\equiv H_1 (k_2/v_2^2)$, 
$h_2\equiv H_2 (k_2/v_2^2)$,
$h_3\equiv H_3 (k_2/v_2^2)$,
$R_M\equiv v_2/\lambda k_2$, 
$\tau \equiv t v_2k_2$, 
use ${\overline {\bfv_2\cdot\curl\bfv_2}}=-f_hk_2v_2^2$.
I assume that ${\overline \beta}_2=q(h_1)v_2/3k_2 $
and ${\overline \beta}_3=q(h_1)v_3/3k_3 $, where $q(h_1)$
is a quenching function for the diffusivity that I take
to be $q(h_1)=(1+ k_1 h_1/k_2)^{-1}$. 
The main points of the present 
study are insensitive to the form of $q$, but  
pinning it down requires more study. 
I also use a  Kolmogorov velocity spectrum  
so that $v_3=(k_2/k_3)^{1/3}v_2$ and  
$\tau_3= (k_2/k_3)^{2/3}\tau_2$.  
The result is 
\beq
\begin{array}{r}
\partial_\tau h_1 = {2\over 3}\left(f_h+h_2+
\left({k_3\over k_2}\right)^{4\over 3}h_3
\right){k_1\over k_2}h_1
-2\left({q(h_1)\over 3}+{q(h_1)\over 3}\left({k_2\over k_3}\right)^{4\over 3}+{1\over R_M}\right)\left({k_1\over k_2}\right)^2h_1,
\end{array}
\label{h1d}
\ee
\beq
\begin{array}{r}
\partial_\tau h_2 = {-2\over 3}\left({k_3\over k_2}\right)^{4\over 3}f_u h_3h_2 -{2\over 3}\left(f_h+h_2\right){k_1\over k_2}h_1
+{2q(h_1)\over 3}\left({k_1\over k_2}\right)^{2}h_1 - 
2\left({q(h_1)\over 3}g_u\left({k_2\over k_3}\right)^{4\over 3}
+{1\over R_M}\right)h_2,
\end{array}
\label{h2d}
\ee
and
\beq
\begin{array}{r}
\partial_\tau h_3 = {2\over 3}\left({k_3\over k_2}\right)^{4\over 3} f_uh_3 h_2 
-{2\over 3}\left({k_3\over k_2}\right)^{4\over 3}\left({k_1\over k_2}\right)h_3h_1
+{2q(h_1)\over 3}\left({k_2\over k_3}\right)^{4\over 3}\left({k_1\over k_2}\right)^{2}h_1 
+ {2q(h_1)\over 3}g_u\left({k_2\over k_3}\right)^{4\over 3}h_2\\
-{2\over R_M}\left({k_3\over k_2}\right)^{2}h_3.
\end{array}
\label{h3d}
\ee
These  equations are to be solved for $h_1$, $h_2$ and 
$h_3$ as a function of time. They can model  
a four-scale nonlinear dynamo theory
in which helicity decays at the resistive scale but
is allowed to transfer between the system scale, the 
forcing scale, and the viscous scale. 
We will compare the solutions with those of the two-scale
system of equations:
\beq
\partial_\tau h_1 = {2\over 3}\left(f_h+h_2\right){k_1\over k_2}h_1
-2\left({q(h_1)\over 3}+{1\over R_M}\right)\left({k_1\over k_2}\right)^2h_1
\label{h1d2}
\ee
\beq
\partial_\tau h_2 = -{2\over 3}\left(f_h+h_2\right){k_1\over k_2}h_1+
{2q(h_1)\over 3}\left({k_1\over k_2}\right)^{2}h_1 - 
{2\over R_M}h_2
\label{h2d2}
\ee
obtained from (\ref{12b}),
(\ref{13b}), and (\ref{14b}) by setting $H_3=0$ and $\beta_3=0$
and then non-dimensionalizing.
Equations (\ref{h1d2})
and (\ref{h2d2}) are essentially the same as those 
solved in Field \& Blackman (2002) and 
which emerged as the specific limit of the strong
damping approximation in the generalized two-scale  theory
of  Blackman \& Field (2002).  Using the doublet (\ref{h1d2}) and
(\ref{h2d2})
of equations rather than the triplet of equations which also includes
$\partial_t\emfb$  cannot capture
possible oscillations in the early time growth of $h_1$.
(Related oscillations were identifed in  simulations by 
Stribling, Matthaeus, Ghosh 1994 
when a uniform field was imposed in their periodic box.)
These can be accounted for only when one allows for a variation in the damping
time in the $\partial\emfb_2$ equation.  
Similarly, the use of the four equations (14-17)
rather  than 6 equations that would result from also including equations for 
$\partial_t\emfb_2$  and $\partial_t\emfb_3$ is similarly restrictive
in the present context.
We leave the necessary  generalizations for future work.

\section{Discussion of Solutions}

The critical helicity for growth of $h_1$
to beat diffusion is $f_h \gsim k_1/k_2$.  
For $f_h$ below this value, $h_1$, $h_2$, and $h_3$ all decay.
This is consistent with numerical simulations of Maron \& Blackman (2002).
Below we focus on the case $f_h=1$ to emphasize the main ideas.
We also assume $g_u=f_u=1$ to simplify the discussion,
and a seed of $h_1\sim 0.001$.
The associated solutions of  
(\ref{h1d}), (\ref{h2d}), and (\ref{h3d}) will be discussed
along with the comparison to solutions of (\ref{h1d2}) and (\ref{h2d2}).  

\subsection{Large scale field growth in the kinematic regime}

By kinematic regime we mean the growth regime in  which $h_1$
grows independently of $R_M$. This is essentially the time for
$h_2$ to approach $-f_h$ (see (\ref{h1d})).  At early times, before
resistivity is important, 
$h_2\sim -h_1$, so we can estimate the duration of the kinematic regime
by determining the time at which $h_1\sim f_h$.
We have, from (\ref{h1d}), 
\beq
t_{kin}\sim 3(k_2/k_1){\rm Ln}(f_h/h_1(0)),
\label{tkin}
\ee
which is $\sim 100$ 
for $f_h=1$, $k_2/k_1=5$ and $h_1(0)=0.001$, as used herein.

A careful look at (\ref{h1d}), (\ref{h2d}), and (\ref{h3d}) 
shows that for  a small seed $h_1$ and 
$h_3(0)\lsim f_h(k_2/k_3)^{4/3}$, the kinetic helicity drives
the growth of $h_1$ and the growth of the oppositely
signed  $h_2$. (If $h_3(0)$ exceeds the above  critical value then
$h_1$ and $h_3$ initially decay until $h_3$ falls below
that value, after which $h_1$ starts to grow.)  
The growth of $h_2$ also supplies, from diffusion,  a growth of 
$h_3$ from the 5th term in (\ref{h3d}).
For $k_2 < k_3 << k_\lambda$, the value of $h_1$ at the end of the 
kinematic phase in the four-scale approach 
saturates at a slightly smaller value of $h_1$ than for the two-scale case. 
The slight difference between the two curves 
can be explained  by ignoring diffusion and dissipation terms 
(those containing $q$ and $R_M$) in 
 (\ref{h1d}), (\ref{h2d}), and (\ref{h3d}) 
and assuming a quasi-steady state. 
Setting time derivatives to zero in 
(\ref{h1d}), (\ref{h2d}), and (\ref{h3d})
then gives a system of three equations:
\beq
(1-k_1/k_2+h_2+(k_3/k_2)^{4/3}h_3) \simeq 0,
\label{s1}
\ee
\beq
h_2h_3=h_1h_3 (k_1/k_2) -h_2 (k_2/k_3)^{8/3}
\sim -h_2 (k_2/k_3)^{8/3}
\label{s2}
\ee
and 
\beq
h_1+h_2+h_3\simeq 0,
\label{s3}
\ee
where we have assumed $h_2 >> h_1k_1/k_2$ anticipating that $h_3$ is small. 
The solution is  
$h_1\simeq{1-k_1/k_2 -(k_2/k_3)^{4/3} +(k_2/k_3)^{8/3}}$,
$h_2\simeq{-1+(k_2/k_3)^{4/3}+k_1/k_2}$,
and $h_3\simeq-(k_2/k_3)^{8/3}$.
Thus no matter what $k_3<<k_\lambda$
is chosen, 
$h_1$ is only slightly depleted from $h_1\sim 1-k_1/k_2$,
the value $h_1$ attains in the 
kinematic regime for the two-scale approach.  
For $k_3\rightarrow k_2$ in 
(\ref{s1}), (\ref{s2}), and (\ref{s3}) $h_1\rightarrow 1-k_1/k_2$.  
Also, for $k_3 >> 1$, $h_1\rightarrow 1-k_1/k_2$.
The correction to $h_1$ is maximized for $k_2/k_3 = 0.6$
at which the deficit is about $25\%$.
In short, $h_1$ $\rightarrow 1$, as in the kinematic regime
here just as in the two-scale approach.
For $k_3=30$ and $R_M= 9000$, these results are seen in the bottom
pair of curves in the upper left plot of Fig. 1, 
where the bold line is the $h_1$ solution of
the two-scale system (\ref{h1d2}) and (\ref{h2d2})
and the thin line is $h_1$ for the four-scale system.
The top set of curves in the same plot  
shows the $k_3=30$ and $R_M=900$ case.
There, $h_3$ first grows as in the previous case, but 
drains more quickly from resistivity.  
Growth of $h_1$ then  more closely mimics the kinematic evolution
of $h_1$ in the two-scale approach at very early times. 

That $h_3\simeq-(k_2/k_3)^{8/3}$ at the end of the kinematic regime
shows that as $k_3$ increases for fixed $k_2$, 
$h_3$ decreases. This is demonstrated in the bottom left plot of Fig. 1 
which shows the  growth of $h_1$ for 
$R_M=100$ and $R_M=2\ts 10^4$ and    
$k_3=160$. There, even the high $R_M$ four-scale solution  
matches the two-scale approach exactly.

\subsection{Rapid migration of the small scale peak to the forcing scale in
the kinematic regime}

An important implication of including the scale $k_3$ can  seen 
by comparing the ratio of the magnetic energy and current helicity at
scales $k_2$ and $k_3$ to those of $k_1$, as a function of time.
This is shown in Figs. 2, 3 and 4 for different
cases. In general,  the current helicity and magnetic energy 
are dominated by the $k_3$ scale at early times. 
These quantities  are related to the magnetic helicity by 
factors of $k^2$ and $k$ respectively.  
The diffusion of $h_2$ into $h_3$ grows enough $h_3$ for
$k_3 h_3$ and $k_3^2 h_3$ to exceed 
the respective values at $k_1$ and $k_2$ initially.  The early
time dominance of the $k_3$ quantities is enhanced when there
is an initial seed of $h_3$, as seen by comparing Figs. 3 and 4.
For a Kolmogorov velocity spectrum however, 
the small scale current helicity and associated magnetic energy 
end up being dominated by the forcing scale before $t_{kin}$: 
as discussed below (\ref{s3}), 
the maximum $h_3$ at the end of the kinematic regime is 
$h_3 \sim -(k_2/k_3)^{8/3}$ so that the ratio of  
current helicity at $k_3$ to that at $k_2$ is 
then  is $(k_3/k_2)^2 (h_3/ h_2) \sim  (k_2/k_3)^{2/3} << 1$,
since $h_2\sim 1$.  

It is interesting to asses whether the times for 
crossover of small scale dominance from $k_3$ to $k_2$ quantities 
depends on magnetic Reynolds number. The answer
is that for large $R_M$ it does not, but for low $R_M$ it
does.  One can be misled if applying the low $R_M$ results
to large $R_M$. 
To see this more explicitly, first note that the location of
the crossovers for the case of Fig. 3 does depend on $R_M$:
The lower the $R_M$ the earlier the crossovers. 
However, for large $R_M$, the location of the crossover
asymptotes to an $R_M$ independent value.
It has been checked that 
the location of the crossovers for  $R_M=9000$ in Fig. 3
is indistinguishable for the location of the cross over for any higher $R_M$,
keeping all other parameters fixed.
This latter result is more directly seen  in the comparisons of the
top and bottom rows of Figs. 3 and 4.   There, the top rows
in both cases correspond to 
a comparison between
$R_M=100$ and $R_M=2\ts 10^4$ for $k_3=160$. The bottom
rows correspond to a comparison between
$R_M=100$ and $R_M=2\ts 10^5$ for $k_3=160$. 
The locations of the cross over for the 
$R_M=2\ts 10^4$ and $R_M=2\ts 10^5$
cases are indistinguishable.

Note that the case of $k_3=160$ and $R_M=100$ for $k_2=5$ is 
the case of $k_3\simeq k_\lambda\simeq k_\nu$ for Kolmogorov turbulence,
in other words,  $Pr_M\simeq 1$. 
This follows because choosing 
$R_M=100$ for $k_3=160$ also corresponds to choosing 
$(k_3/k_2)=R^{3/4}\simeq 100$, 
where $R$ is the hydrodynamic
Reynolds number and where $k_3$ is taken to be the 
viscous wavenumber.  The power $3/4$ follows from the Kolmogorov spectrum. 
This can be distinguished from any $R_M >> 100$ for fixed $k_3$, 
which, in turn, represents  $Pr_M >> 1$.  

Thus, restating the results of this section  in terms of $Pr_M$, 
we can say that for the $Pr_M = 1$ case 
the crossover time for the current helicity 
at $k_3$ to deplete below that at  $k_2$ is earlier than 
for the $Pr_M >>1$ case  in which $k_3 << k_\lambda$.
This is  expected,
since the resistivity is not effective at dissipating
$k_3$ for $Pr_M >>1$. At large $R_M$ for fixed $k_3=k_\nu$
(or $Pr_M >>1 $), 
the crossover is the result of the dynamical depletion 
of $h_3$ from the 
first two terms on the right 
of (\ref{h3d}), and becomes independent of $R_M$ or $Pr_M$.

\subsection{Saturation: the doubly maximal inverse transfer state}

The growth term for $h_3$ at early times is the fifth term in
(\ref{h3d}), but eventually this is offset by the
fourth term, after which the only remaining terms for $h_3$
are resistive and inverse transfer loss terms.
Thus, as we have seen,  $h_3$ eventually decays, whilst $h_2$  
continues to grow.  The latter takes over the role
of compensating negative helicity, and the saturation proceeds
exactly  in the two-scale approach:  positive helicity at $h_1$
and negative helicity $h_2$ at the forcing scale.  This is demonstrated
in the right column of Fig. 1 where the 
late time evolution of  the two-scale and four-scale approaches
are seen to be indistinguishable for all $R_M$.
  Accordingly, at late times, the current
helicity at $k_2$ asymptotes to equal that at $k_1$, just as in the two-scale
approach (Brandenburg 2001; Field \& Blackman 2002; Blackman \& Field 2002).

The features of the four-scale model discussed herein 
can be used as an aid to understand  
the full saturated spectrum of the  helical dynamo 
both above and below the forcing scale.
Consider $k_L\le k_1< k_f\le k_2 < k_3 \le k_\nu$, 
where $k_L$ is the wavenumber of the largest scale available,
$k_f$ is the forcing scale and $k_\nu$ is the viscous scale
and we allow $k_1,k_2,k_3$ to take on intermediate values.
The time for the spectrum to evolve depends on the ratios 
$k_L/k_f$ and $k_f/k_\nu$, 
but the qualitative evolution can be understood simply.
At early times, a seed spectrum forced with negative kinetic helicity
with $f_h > k_1/k_f$, can grow positive magnetic
helicity at $k_1$.  (For $f_h=1$ the maximum growth of the large scale
magnetic  helicity initially  occurs at $k_1 \simeq k_2/2$.) 
The magnetic helicity will inverse transfer from $k_1 > k_L$
to successively smaller $k$, eventually
all the way to $k_L$. The $k_L$ will continue to accumulate
the bulk of the positive magnetic helicity until saturation.
This happens by direct analogy to the sub-forcing scale
dynamics discussed earlier. There the negative magnetic helicity initially 
grows fastest on the scale $k_\nu$ before loss terms eventually
win on that scale, and the negative magnetic helicity
transfers upward to larger scales, eventually 
reaching the forcing scale where the bulk of the negative magnetic helicity
collects. {\it The positive and negative magnetic helicities
thus migrate to their largest available scales.}
The negative helicity  cannot grow at any $k$ significantly less than $k_f$
since the growth rate for positive magnetic helicity is positive
for lower $k$.
The overall picture just outlined for saturation seems
to be consistent with simulations of Brandenburg (2001) and Maron \&
Blackman (2002).

Note that the larger the scale, the less important
the resistive terms in draining the helicities
from those scales.  The dominant loss terms
for $k<< k_\lambda$ are the inverse transfer
terms, not resistive terms. The threshold magnetic helicity
at a given scale for the inverse transfer loss  
to dominate its gain from diffusion of magnetic helicity
from larger scales, increases with decreasing $k$.
Thus, larger scales take longer to inverse
transfer their magnetic helicity.
For the positive helicity above the forcing scale,
this means that the $k_L$ scale takes the longest to grow to its
maximum, and for the negative helicity, $k_f$ takes the longest
to grow to its maximum.
The final state can be described as a state of 
``doubly maximal inverse transfer.''


\subsection{Implications and comparison to fractionally helical dynamo spectra}

Maron \& Blackman (2002) showed that as $f_h$ exceeds $k_L/k_f$ and approaches
$1$ for $Pr_M > 1$, the saturated magnetic energy spectrum changes from peaking
on the resistive scale to peaking on two scales, with one peak at the forcing
scale and one at the system scale.  (Their $f_h=1$ results matched 
Brandenburg 2001.)  For $f_h > k_L/k_f$, 
 $f_h(k_L/k_f)$ times the equipartition energy ends up in magnetic energy 
at the  the $k_1$ scale after the kinematic regime.
At saturation, $ f_h k_f/k_L$  of the equipartition energy
ends up at $k_L$. 

The fraction of small scale magnetic energy associated with the
helicity dynamics is also determined by $f_h$.
Whatever the dynamics of the nonhelical magnetic energy spectrum, 
the helical fraction of the magnetic 
spectrum seems to be  explicable by the independent 
dynamics discussed above, suggesting that the helical and
nonhelical parts of the spectrum are rather decoupled.
Recall that the nonhelical magnetic energies on the $k_3$ and 
$k_\lambda$ scales do not enter the theory, only that on 
the $k_2$ scale (see the discussion
between equations (\ref{14bb}) and  (\ref{h1d})).
In the present theory, the expected helical fraction of the small scale 
magnetic energy that follows the 
helicity dynamo dynamics is given simply by $f_h$.
This follows because the small scale current helicity in saturation 
is  $f_h v^2 k_2 \sim \lb \bfb\cdot \curl \bfb\rb$ and 
since $\lb\bfb^2\rb \sim \lb\bfv^2\rb$ in saturation,
we have the associated helical magnetic energy fraction $=f_h$.
If we include the nonhelical component of the magnetic energy,
then $f_h$ is a lower limit to the fraction of 
the total small scale magnetic energy that winds up at the forcing scale.

\section{Conclusions}

A four scale nonlinear magnetic dynamo theory
was presented using the approximation that correlations of
mixed scales are assumed to vanish.  
The goal was to develop a simple theory that sheds light
on the evolution of the full magnetic spectrum for the helical dynamo
and show how the presence of kinetic helicity 
influences the spectrum both above and below the dominant 
scale of the turbulent kinetic energy.
The velocity spectrum was assumed
to be Kolmogorov, with kinetic helicity input only at the forcing scale.  
The results of the simple theory are consistent
with existing numerical simulations and make additional predictions which
can be tested. The theory includes the dynamical backreaction
of the growing magnetic field on the turbulence driving the field
growth.  

The growth of the large scale field in the four scale theory
is consistent with that predicted in the  
two-scale theory (Field \& Blackman (2002); Brandenburg \& Blackman (2002);
Blackman \& Field  (2002)) at late times but previous
simulations do not have enough resolution to test
the dynamics of the kinematic regime and so higher
resolution simulations will be needed. 
The four-scale theory herein predicts  that as the large 
helical scale field grows, the small scale helical field of the opposite
sign and its associated current
and magnetic helicity will first grow at the smallest
scales where both $\bfv$ and $\bfb$ are finite (for $Pr_\ge 1$, this
is the viscous scale).  This cannot be seen in a two-scale approach, since 
there the system scale and forcing scales are the only ones present.
The source of helical field for the very 
smallest scales is diffusion from above. 
These very small (viscous) scale helical fields drain by
inverse transfer, and  the helical magnetic energy below the forcing scale
peaks at successively larger 
scales, arriving at the forcing scale before $t_{kin}$.  
If the viscous scale is much
larger than the resistive scale, then resistivity plays
little role in this process;  the process is 
independent of $R_M$ for large $R_M$. 
At $t_{kin}$ the growth of both the large and small scale 
helical fields proceed exactly as in the two-scale theory:
large scale current and magnetic helicities are primarily carried
by $k_1$, and the oppositely signed small scale quantities at scale $k_2$.

That high resolution simulations are  needed to test the kinematic
regime also means that they are need for testing the assumption
made above Eqn. (7).  If future simulations for $Pr_M >>1$ 
show that the current helicity 
piles up at the resistive scale rather than the viscous scale at 
very early times, this would suggest
that the approximation that correlations of mixed 
scales vanish would have to be revised, since it is only such
mixed correlations which can grow helicity on the resistive scale.
In the present, four-scale theory, the helical magnetic energy 
on the resistive scale decays, and the nonhelical component is decoupled
from the helical component.   I emphasize that the model
presented herein is a simplified theory, meant to guide
physical understanding of helical MHD turbulence. It must 
be subject to further testing.

The larger the fractional kinetic helicity $f_h$, 
the more the overall magnetic spectrum behaves like its helical component.
If the nonhelical component prefers to pile-up at small
scales, the helical component will provide at least some fraction that
wants to be doubly peaked at the forcing and system scales. 
It is thus important to emphasize that 
that driving turbulence with kinetic helicity 
on a scale $k_2$ affects  the shape
of the overall helical magnetic energy spectrum and thus that of 
the overall magnetic energy spectrum above and below
the forcing scale.

\ni 
Many thanks to G. Field, A. Brandenburg, B. Chandran, H. Ji, R. Kulsrud 
and J. Maron for  discussions. DOE grant DE-FG02-00ER54600 is acknowledged.
Thanks to the Department of Astrophysical 
Sciences and the Plasma Physics Lab at Princeton for hospitality 
during a sabbatical.

\vfill
\eject

\vspace{-.1cm} \hbox to \hsize{\epsfxsize7.5cm
\epsffile{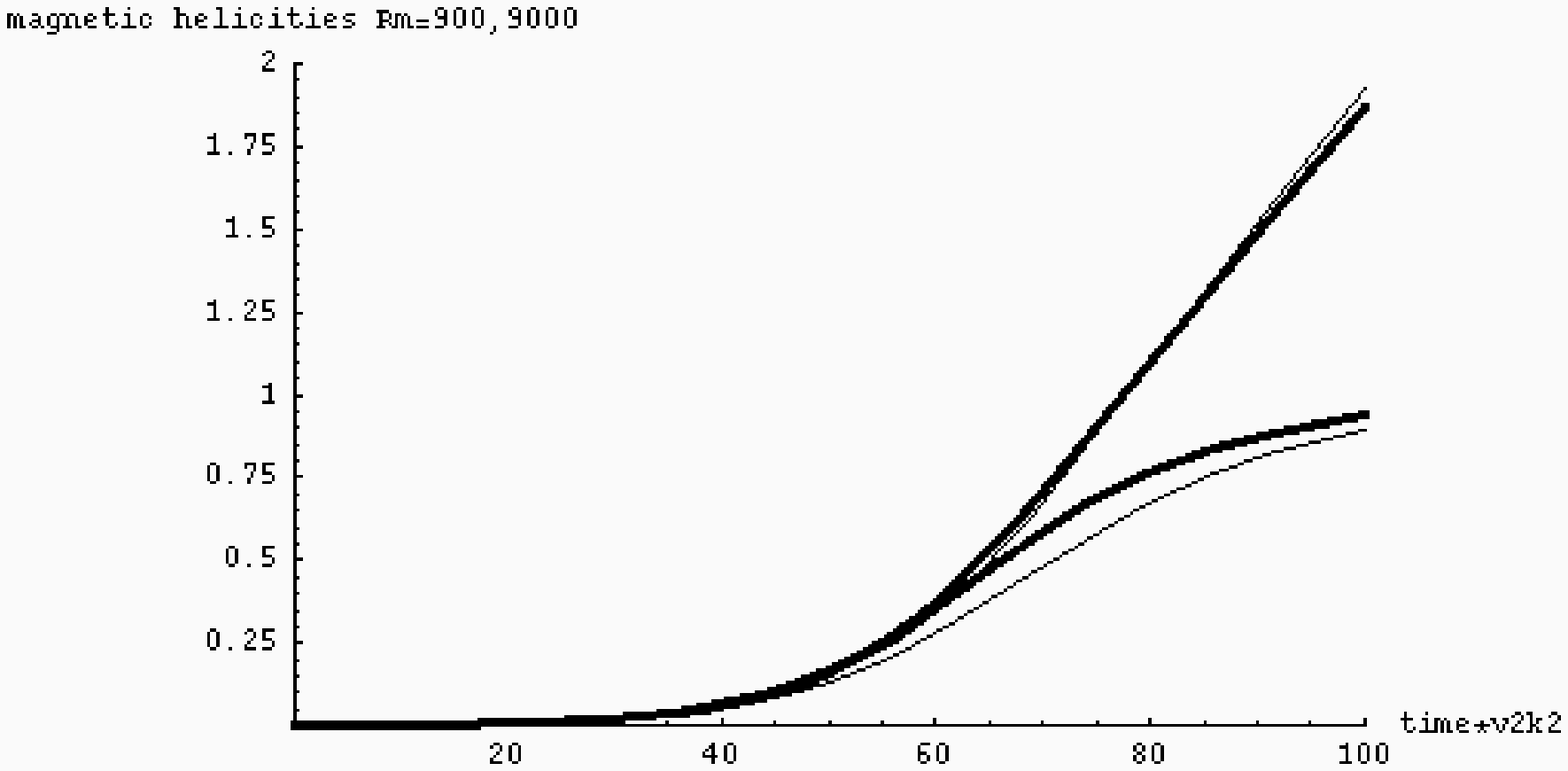} \epsfxsize7.5cm \epsffile{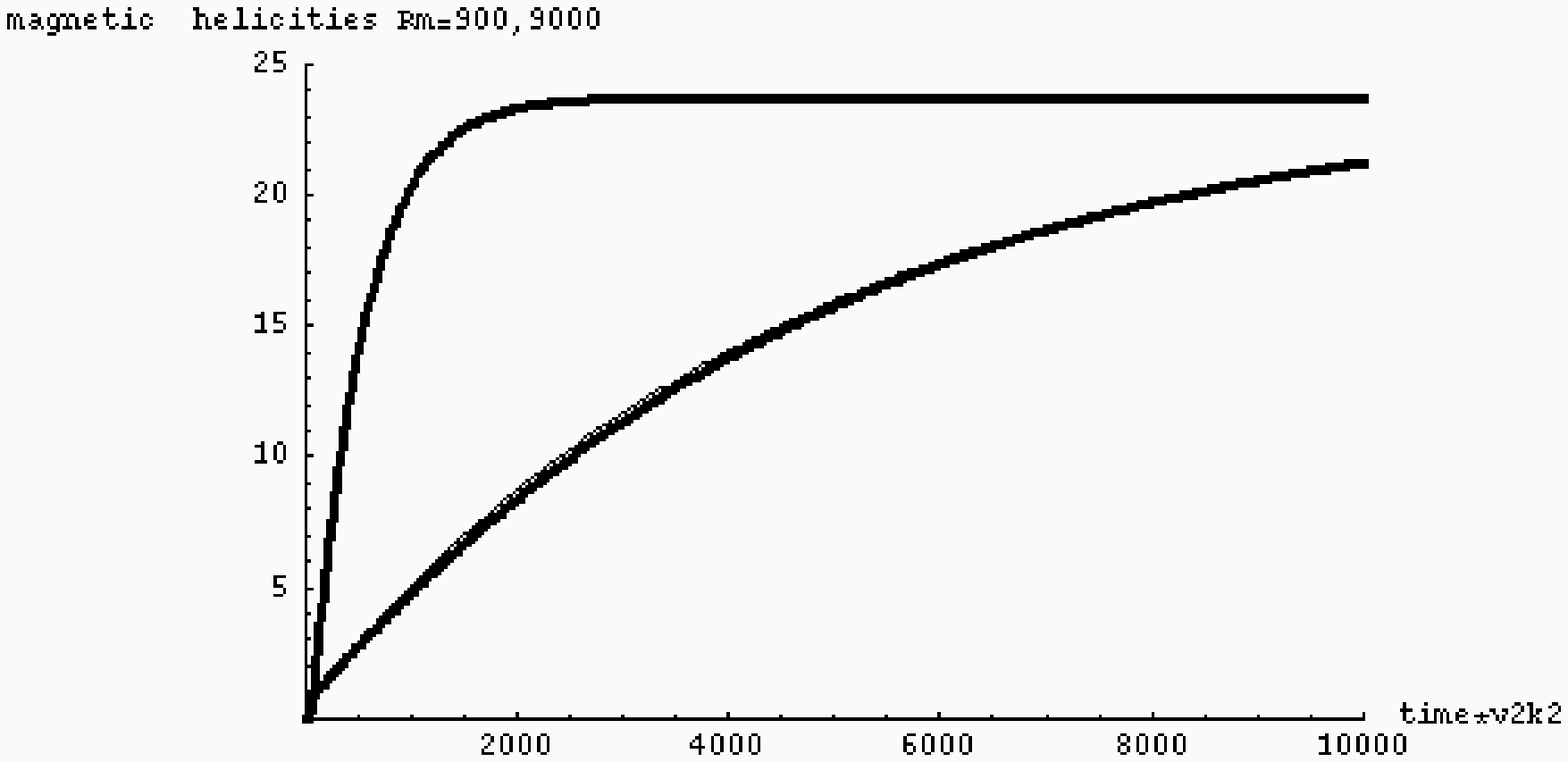}}
\vspace{-.1cm} \hbox to \hsize{\epsfxsize7.5cm
\epsfxsize7.5cm \epsffile{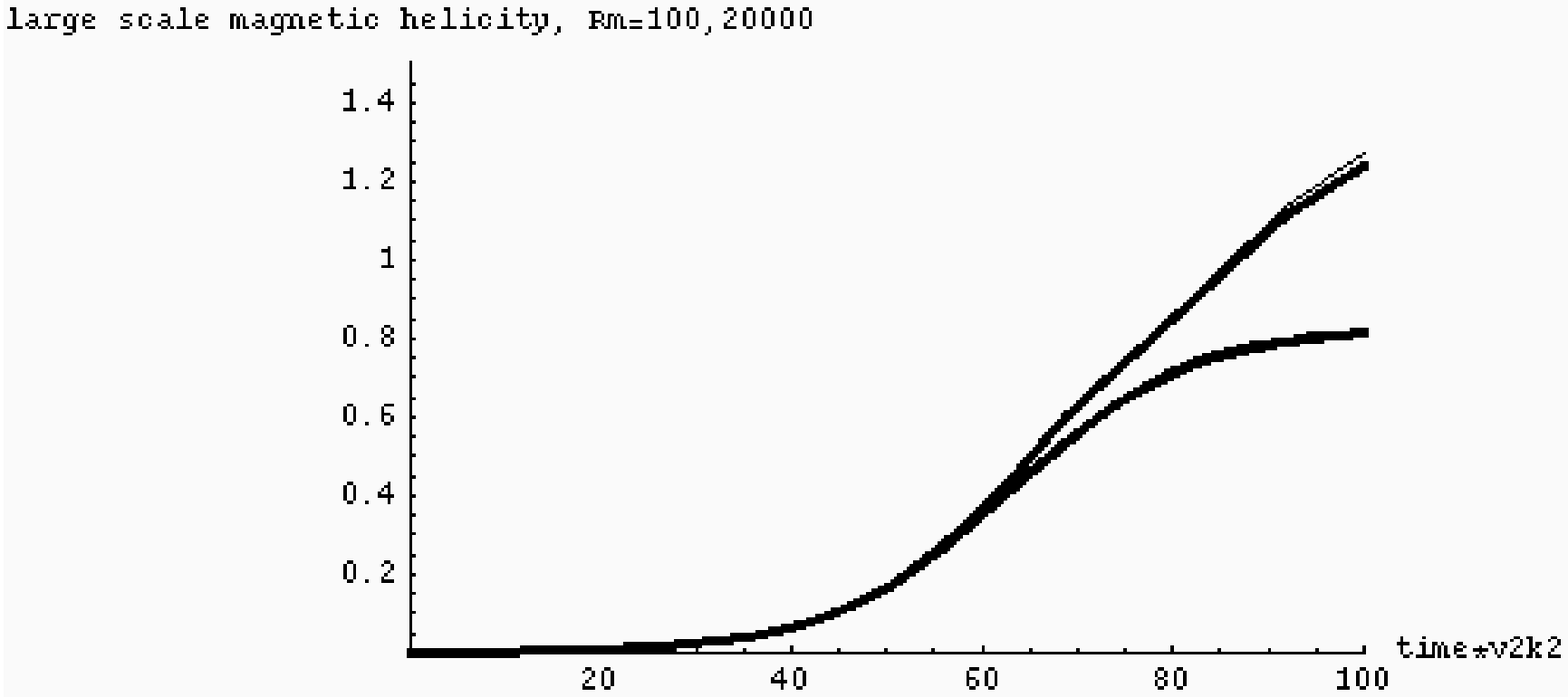} 
\epsfxsize7.5cm \epsffile{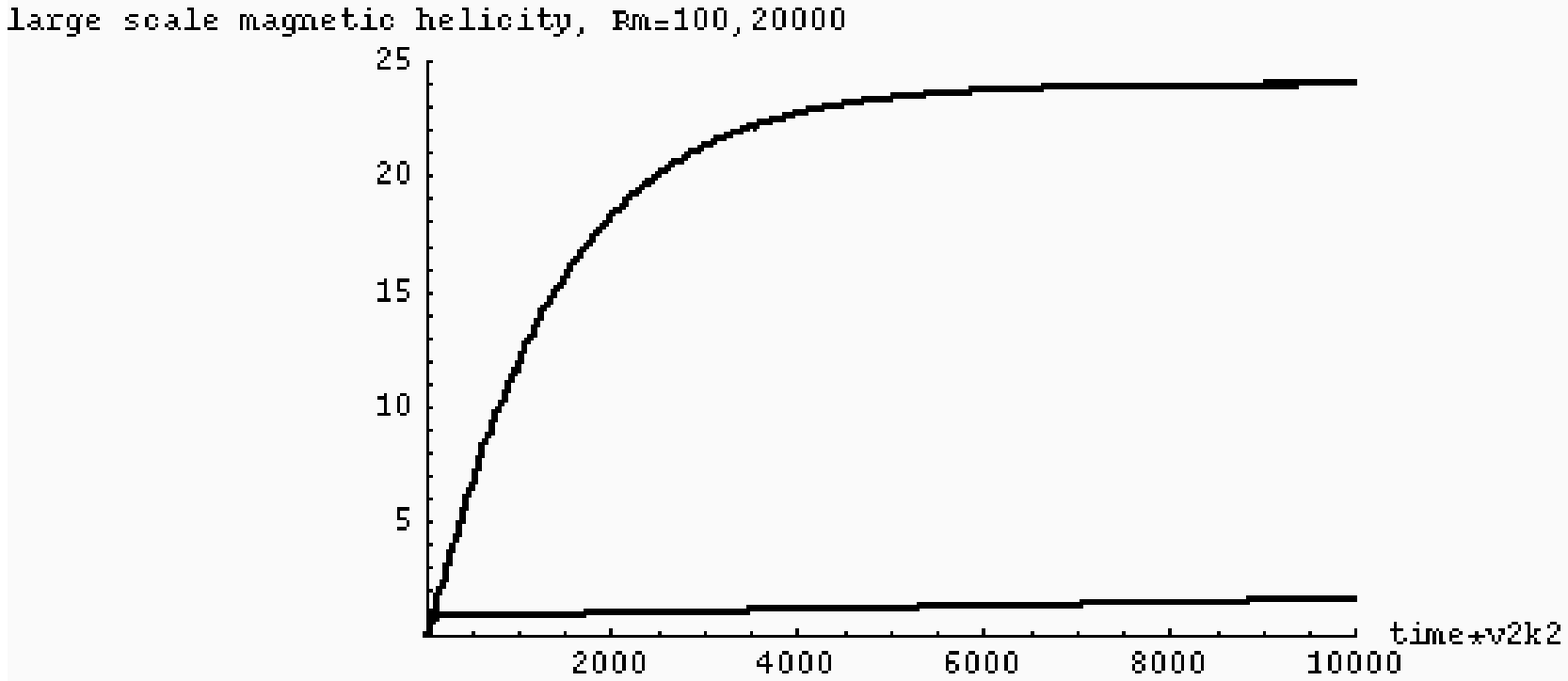}}
\noindent{Figure 1:  The magnetic helicity, $h_1$, for the
four-scale approach (thin lines) and 
the two-scale
approach (thick lines). The left column of plots 
is for early times and the right column
is  for a broader time range. 
For the top row, $k_1=1$, $k_2=5$, $k_3=30$, 
and $R_M = 900$ (top pair of curves) $R_M = 9000$ (bottom pair of curves). 
For the bottom row, $k_1=1$, $k_2=5$, $k_3=160$,  and 
$R_M = 100$ (top pair of curves)
$R_M = 2\ts 10^4$ (bottom pair of curves). Note that the two-scale
and four-scale approaches are largely indistinguishable in all but
the upper left plot (see text). 
(The high $R_M$ curve in the lower right 
will eventually saturate at the same value $h_1= (k_2/k_1)^2$ 
as the low $R_M$ case but at much later times and thus undershoots
the top curve for the plotted time range.)} 

\vspace{-.1cm} \hbox to \hsize{ \hfill 
\epsfxsize7.5cm \epsffile{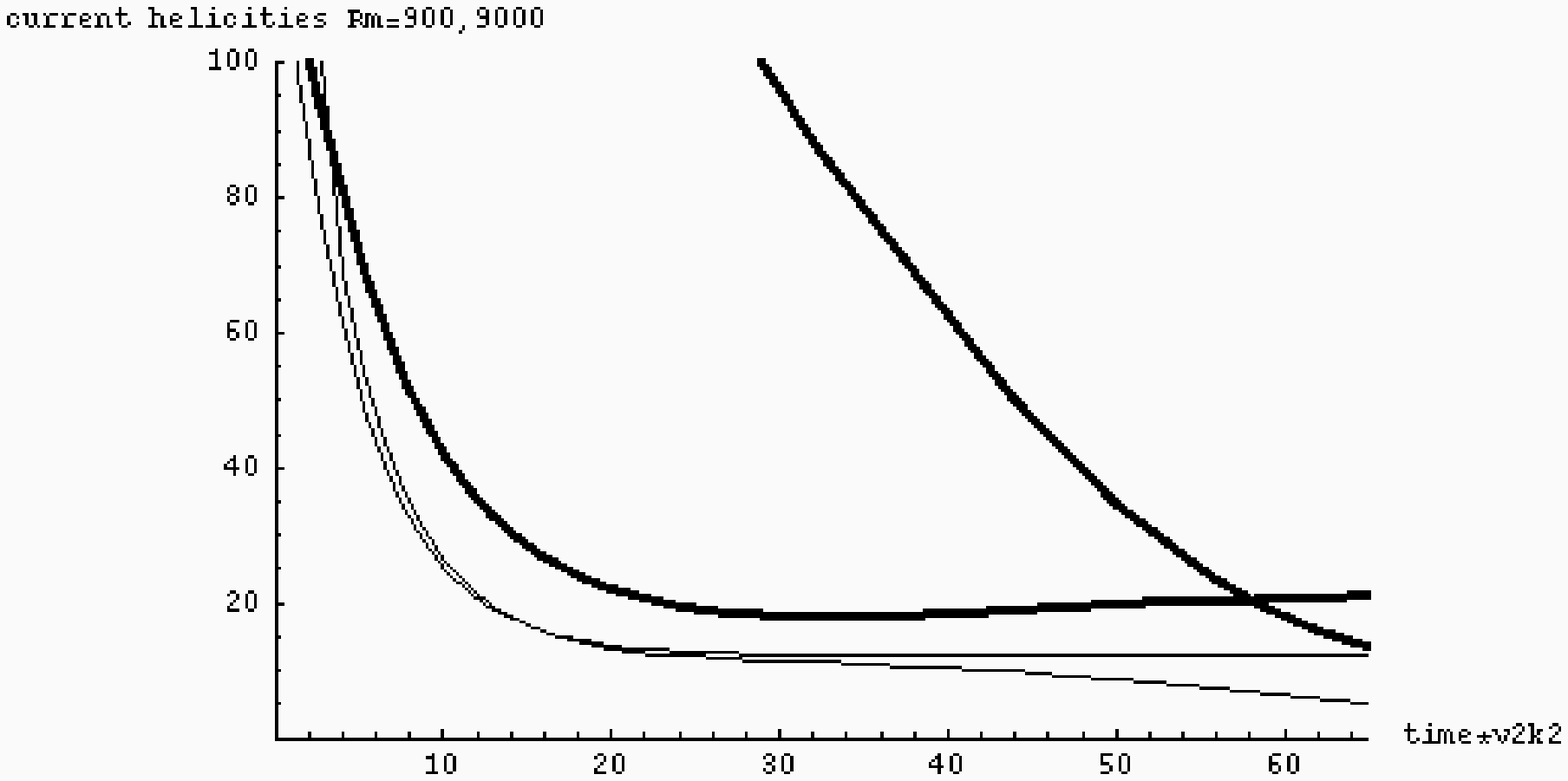} \epsfxsize7.5cm 
\epsffile{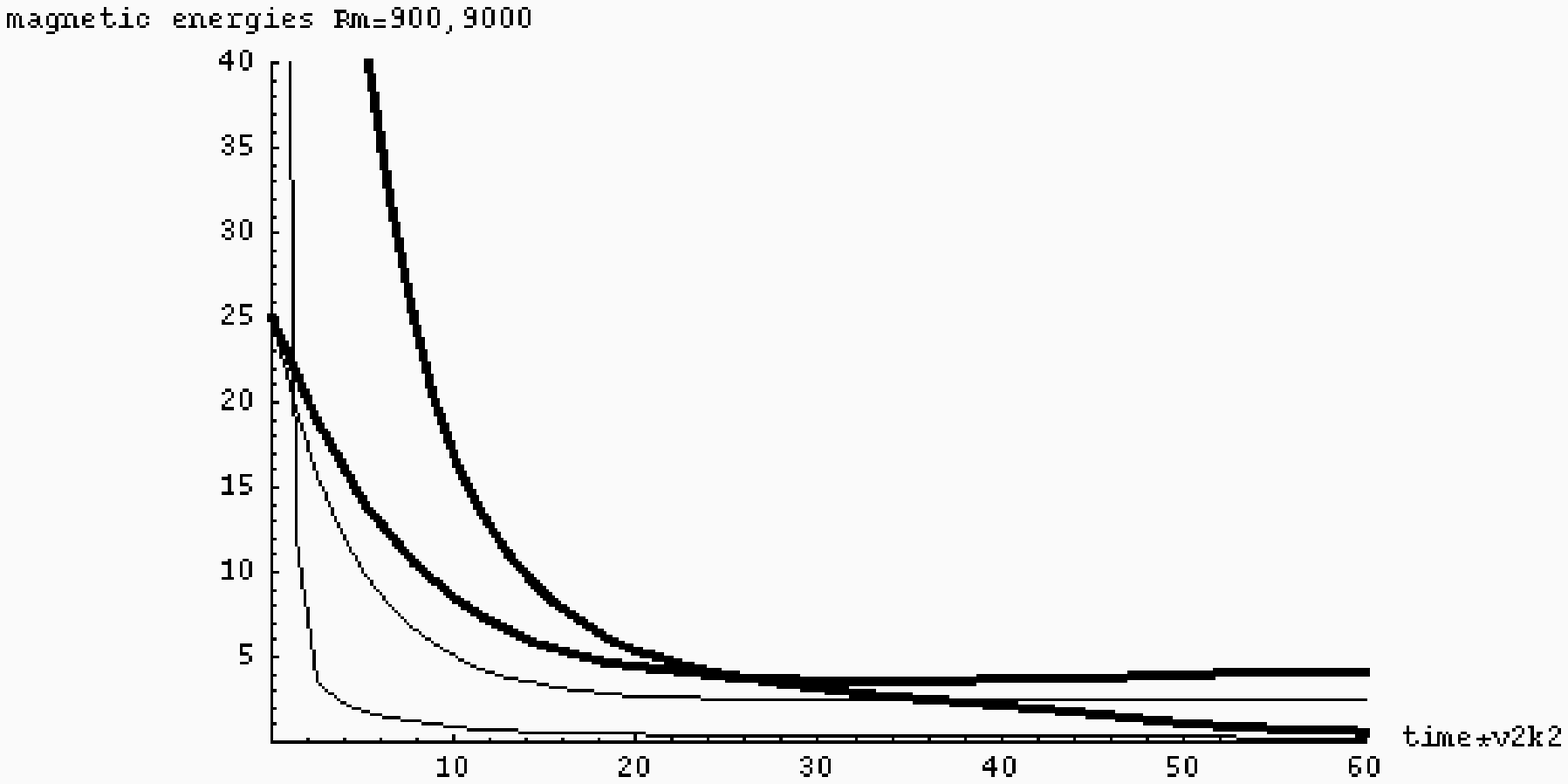} 
\hfill}
\noindent{Figure 2: 
The absolute magnitude of the current helicity
ratios $|k_2^2 h_2/k_1^2 h_1|$ and $|k_3^2 h_2/k_1^2 h_1|$, and 
helical magnetic energy ratios  
$|k_2 h_2/k_1 h_1|$ and $|k_2 h_2/k_1 h_1|$   assuming
an initial seed of $h_2(0)=h_3(0)=-h_1(0)/2=0.0005$.
Here $R_M = 900$ (thin lined curves)
$R_M = 9000$ (thick lined curves)
for $k_1=1$, $k_2=5$, $k_3=30$. 
In each plot, and for both
values of $R_M$, the quantities at $k_3$ dominate at early
times, and then become subdominant to the values at 
$k_2$ at later times.  The location of the crossover
is earlier for $R_M=900$ because the resistive scale is closer
to the scale $k_3$; resistivity is more effective
at early times in draining the  
current helicity and magnetic energy at $k_3$ than in the $R_M=9000$
case.}

\vspace{-.1cm} \hbox to \hsize{ \hfill 
\epsfxsize7.5cm \epsffile{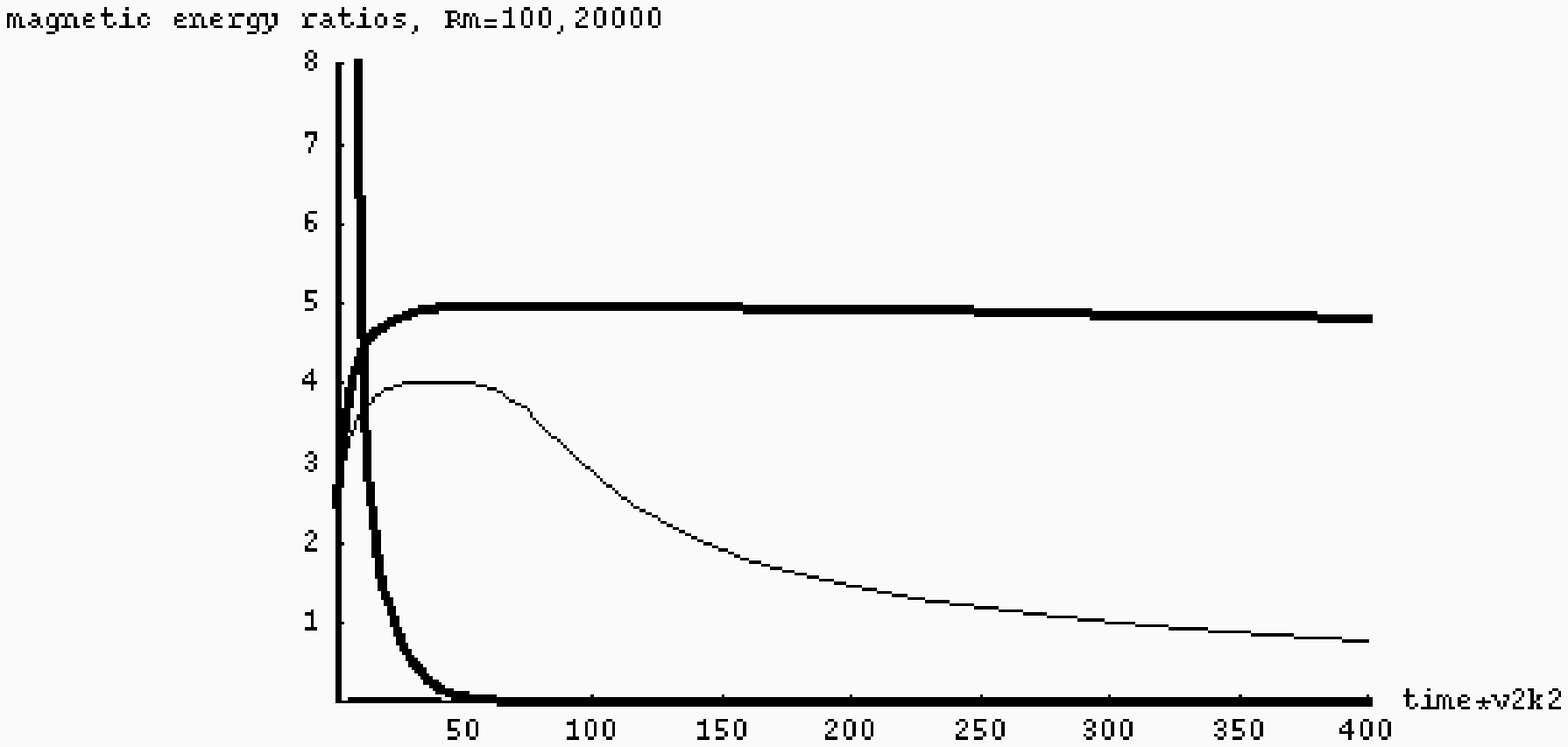} \epsfxsize7.5cm 
\epsffile{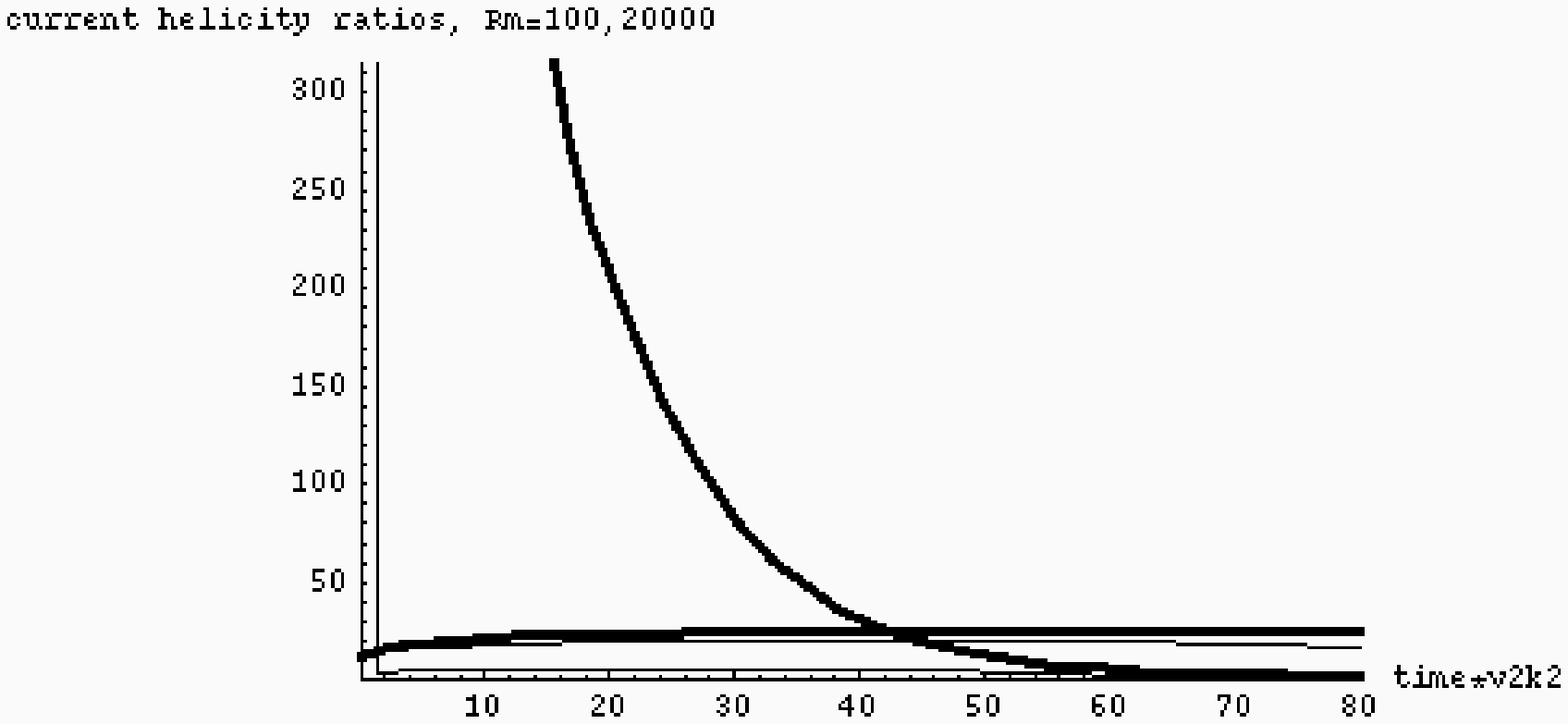} 
 \hfill }  
\vspace{-.1cm} \hbox to \hsize{ \hfill 
\epsfxsize7.5cm \epsffile{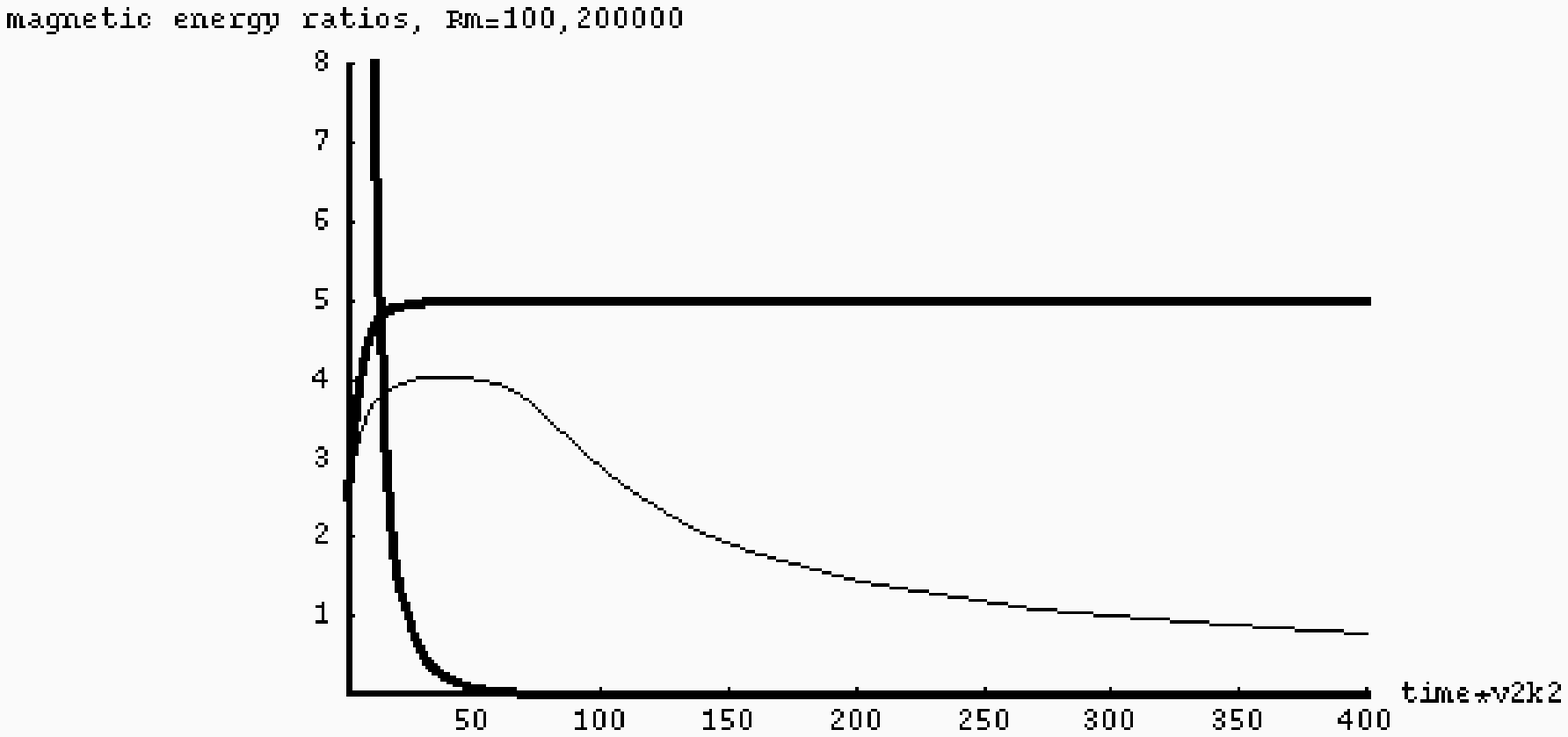} \epsfxsize7.5cm 
\epsffile{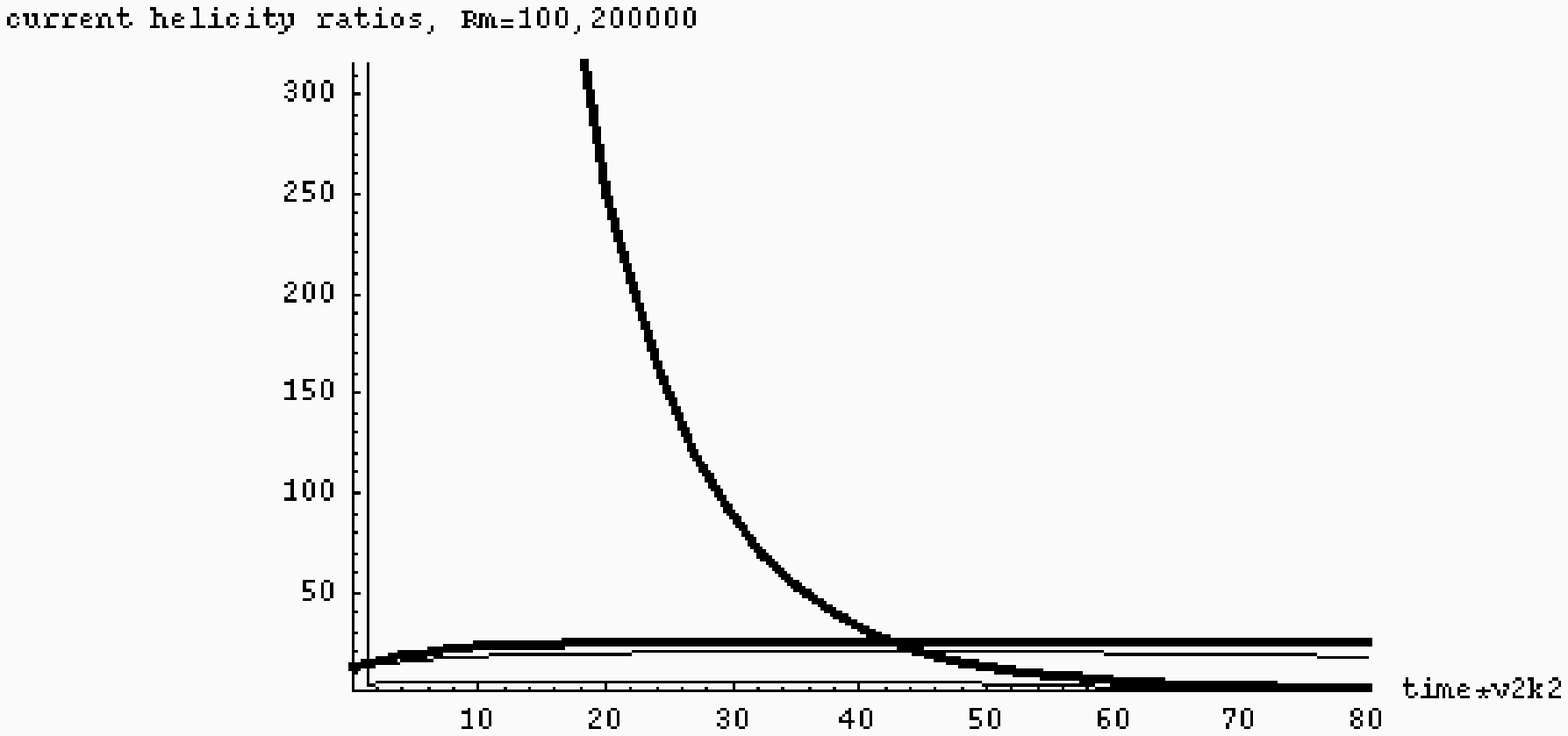} 
 \hfill }  
\noindent{Figure 3: 
The absolute magnitude of the current helicity
ratios $|k_2^2 h_2/k_1^2 h_1|$ and $|k_3^2 h_2/k_1^2 h_1|$,  and 
helical magnetic energy ratios  
$|k_2 h_2/k_1 h_1|$ and $|k_2 h_2/k_1 h_1|$   assuming
an initial seed of $h_2(0)=h_3(0)=-h_1(0)/2=0.0005.$ 
for $k_1=1$, $k_2=5$, $k_3=160$. Top row has plots for 
$R_M = 100$ (thin lined curves) and 
$R_M = 2\ts 10^4$ (thick lined curves), 
and bottom row has plots for $R_M = 100$ (thin lined curves)
and $R_M = 2\ts 10^5$ (thick lined curves).
In each plot,  the quantities at $k_3$ dominate at early
times, but then become subdominant to the values at 
$k_2$ at later times.  
The magnetic energy plot is shown for a broader
time range.
The location of the crossover
for the large $R_M$ cases of $2\ts 10^4$ and $2\ts 10^5$
is independent of $R_M$: the crossovers 
occur for the thick pairs of lines at the same 
time in the top and bottom rows. 
The crossover occurs much earlier for $R_M=100$ because
the resistive wavenumber is closser
to $k_3$ and thus resistivity is more effective
at early times in draining the  
current helicity and magnetic energy at $k_3$ than in the larger
$R_M$ cases. For the current helicity, the cross over for $R_M=100$
occurs so early that it is not visible on the graph.
Before the end of the kinematic regime ($t\lsim 100$),
the crossovers are complete and $k_2$ emerges as the dominant small scale. 
Since $k_3=160$ here, $R_M=100$ corresponds to $Pr_M\simeq 1$
}

\vspace{-.1cm} \hbox to \hsize{ \hfill 
\epsfxsize7.5cm \epsffile{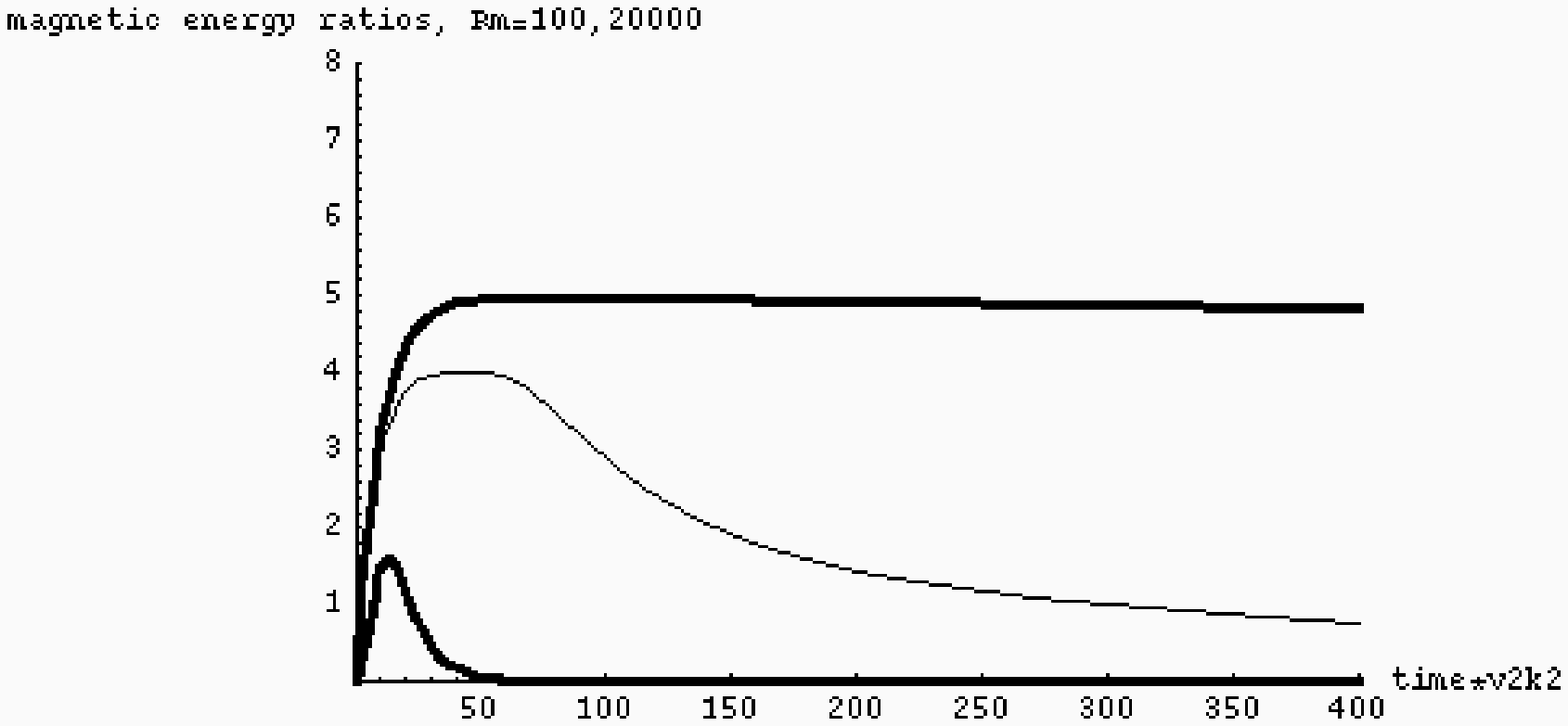} \epsfxsize7.5cm 
\epsffile{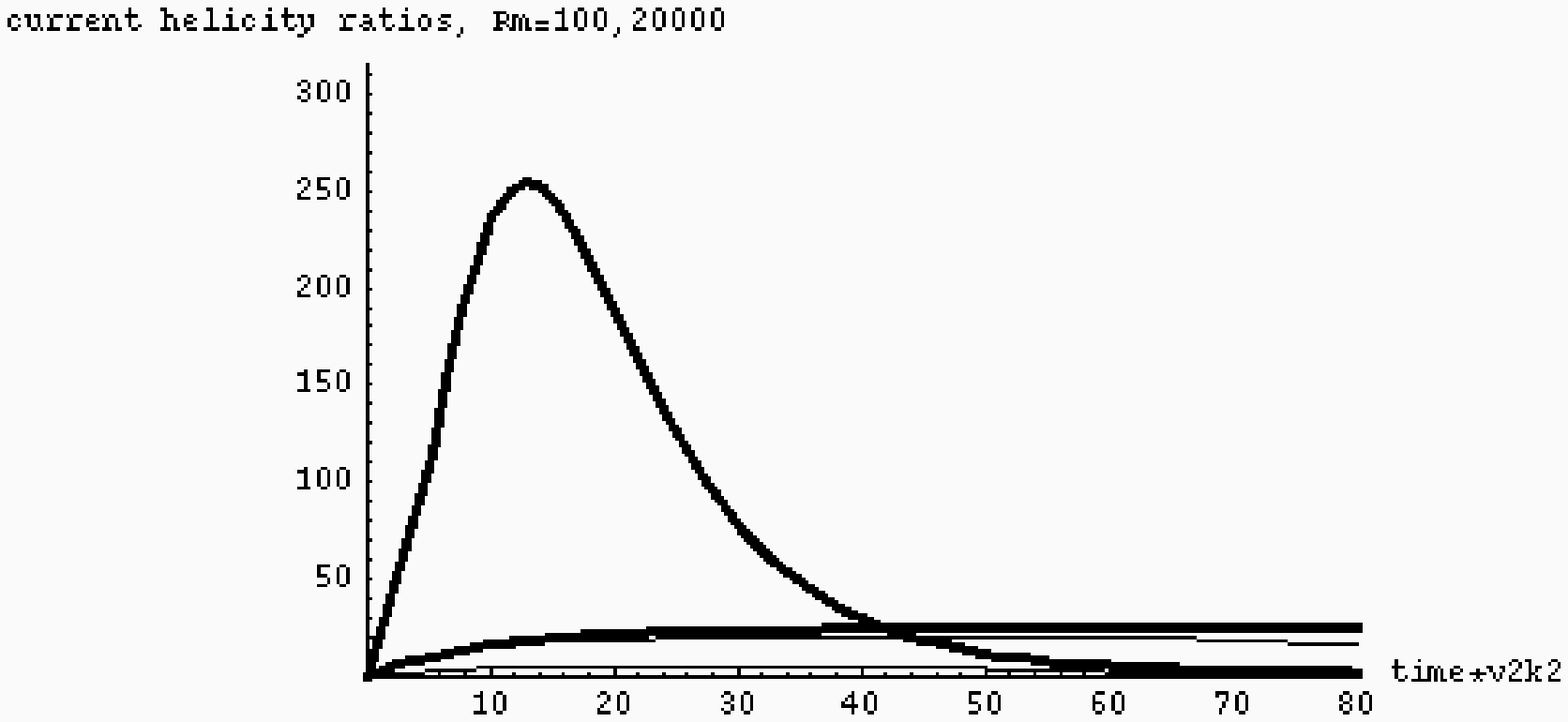} 
 \hfill }  
\vspace{-.1cm} \hbox to \hsize{ \hfill 
\epsfxsize7.5cm \epsffile{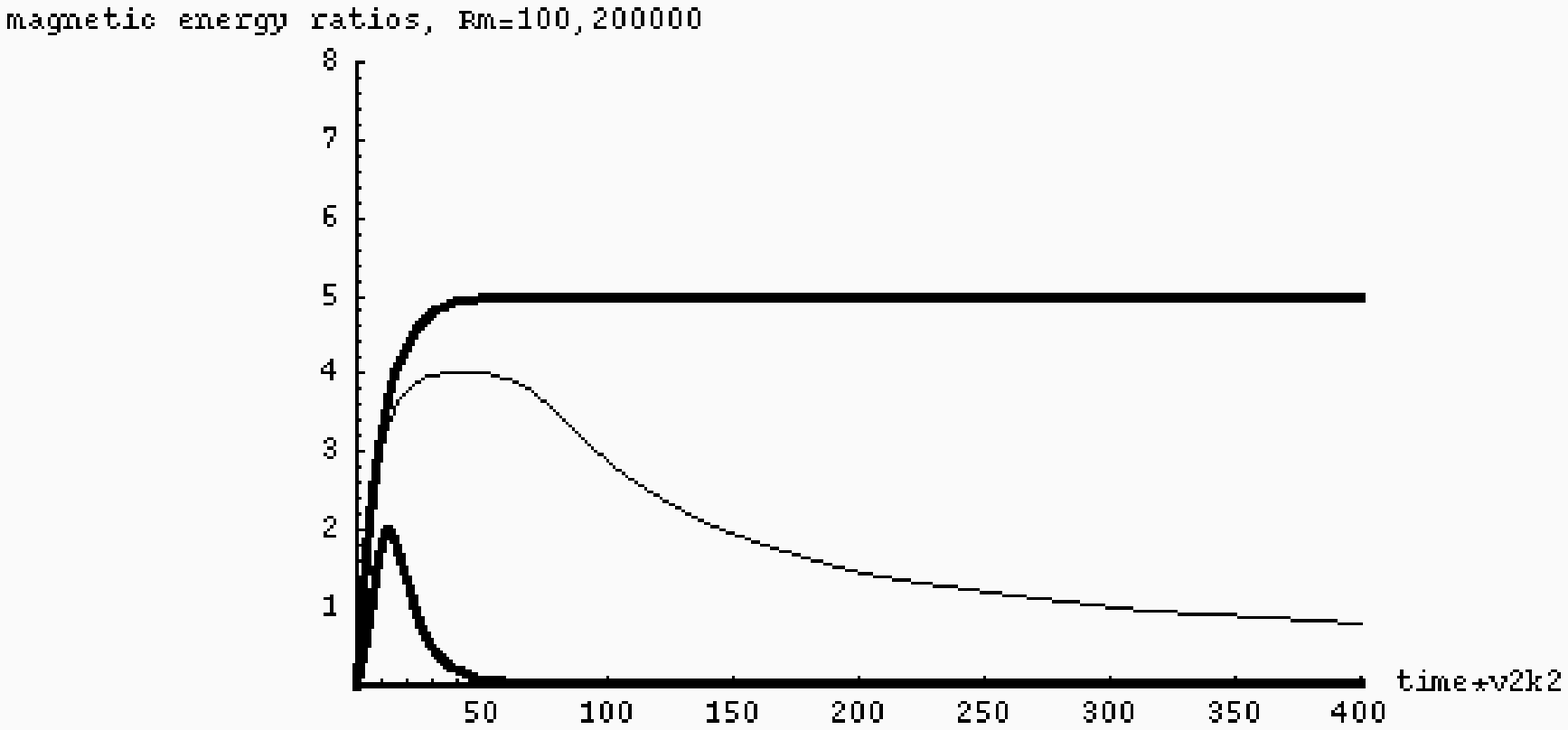} \epsfxsize7.5cm 
\epsffile{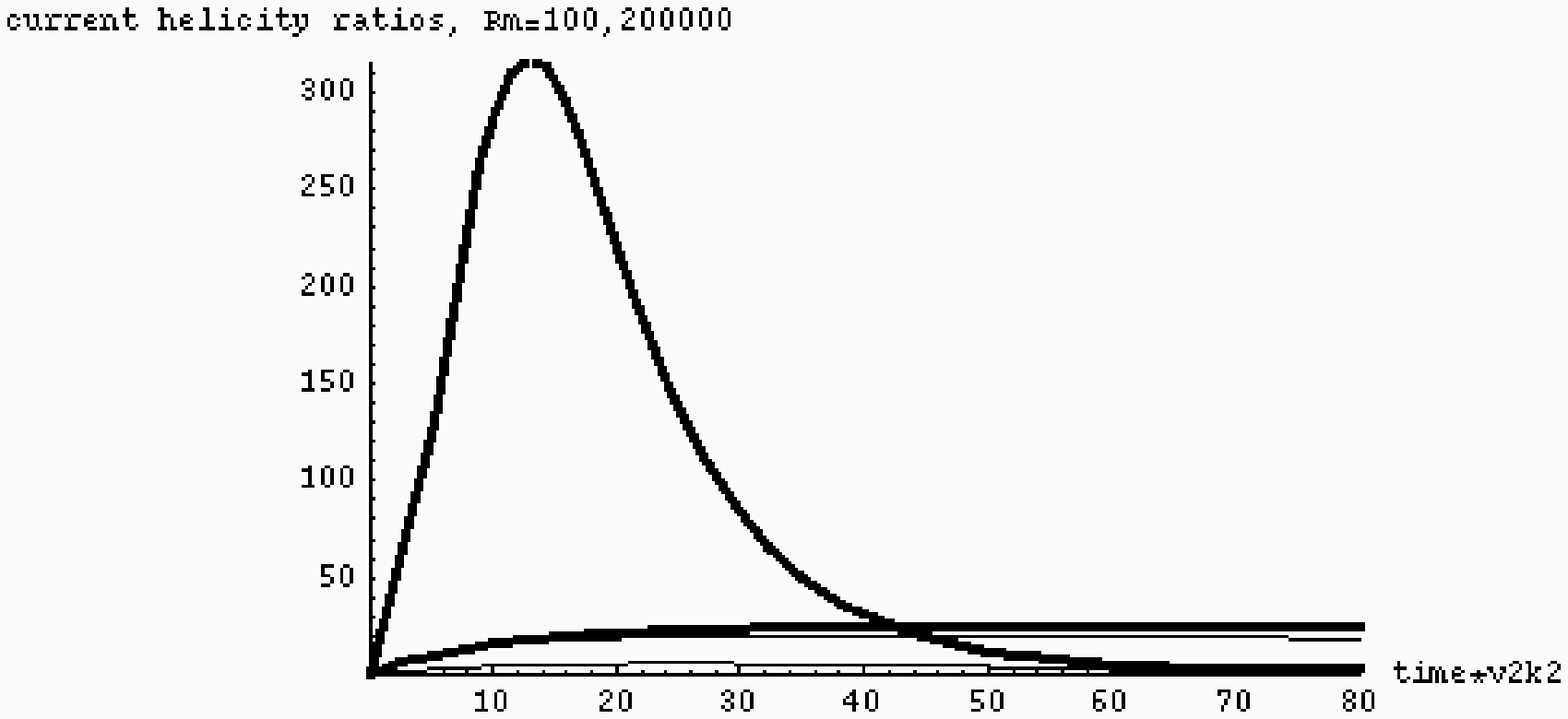} 
 \hfill }  

\noindent{ Figure 4: 
Same as Figure 3, but with $h_1(0)=0.001$
and $h_2(0)=h_3(0)=0$.  Note the initial rise of
the current helicity and magnetic energy at $k_3$ but then again the dominance
of the $k_2$ quantities before the end of the kinematic regime
$t_{kin}=100$}

\end{document}